\documentclass[aps,prl,floats,twocolumn,showpacs]{revtex4}
\usepackage{graphicx}

\begin{document}
\title{Signatures of Spin in the $\nu =1/3$ Fractional Quantum Hall Effect}
\author{A. F. Dethlefsen$^{1}$, E. Mariani$^{2}$, H.-P. Tranitz$^{3}$, W. Wegscheider$^{3}$, and R. J. Haug$^{1}$
    \vspace{1mm}}
\affiliation{$^1$Institut f\"ur Festk\"orperphysik, Universit\"at Hannover, 30167 Hannover, Germany \\
$^2$Department of Condensed Matter Physics, The Weizmann Institute of
Science, 76100 Rehovot, Israel
\\
$^3$Institut f\"ur Angewandte und Experimentelle Physik Universit\"at Regensburg, 93040 Regensburg, Germany
\vspace{0mm}}
\date{\today}
\begin{abstract}The activation gap $\Delta$ of the fractional quantum Hall state at constant filling $\nu =1/3$ is
measured in a \emph{wide} range of perpendicular magnetic field $B$.
Despite the full spin polarization of the incompressible ground state, we
observe a sharp crossover between a low-field linear dependence of
$\Delta$ on $B$ associated to spin texture excitations and a Coulomb-like
behavior at large $B$. From the global gap-reduction we get information
about the mobility edges in the fractional quantum Hall regime.
\end{abstract}
\maketitle
\newcommand{\omecst}{\omega_{\mathrm{c}}^{*}}
\newcommand{\omec}{\omega_{\mathrm{c}}^{}}
\newcommand{\muB}{\mu_{\tiny{\mathrm{B}}}^{}}

In recent years the increased mobility of two-dimensional electron systems
(2DES) has allowed for the experimental investigation of the fractional
quantum Hall effect (FQHE) \cite{Tsui,Laughlin} at relatively small
magnetic fields. In this regime the interplay between interactions and the
electronic spin yields interesting properties of either the ground state
or the excitation spectrum of the system. Quantum phase transitions
between differently spin-polarized ground states have been predicted and
experimentally observed for several FQH states while varying the
perpendicular magnetic field $B$
\cite{Halperin83,Chak84,Chak86,Haug87,ParkJain98,Eisenstein89,Engel92,Du95,Kukushkin99,Kukushkin00}.

In parallel, the theoretical understanding of the FQHE has been considerably deepened by the introduction of Composite
Fermions (CF)
\cite{Jain89}, quasiparticles made of one electron and two magnetic flux quanta. The correlated many-electron problem in
high magnetic fields can be interpreted in terms of weakly interacting CF in a smaller effective field, offering a
unified theory of the compressible and incompressible electronic FQH states. The former are mapped onto CF Fermi liquids
\cite{HLR93}, while the latter are viewed as the integer QHE of CF.

One of the main experimental signatures associated to the incompressible
FQH states is a thermally activated longitudinal resistivity,
$\rho_{xx}^{}\propto \exp [-\Delta /2T]$, with the associated finite
activation gap $\Delta$. An accurate determination of $\Delta$ is
therefore crucial in order to test the predictions of the CF theory on the
spectrum of incompressible FQH states and to extract the quasiparticle
effective parameters. Early measurements of $\Delta$ vs $B$ revealed the
importance of disorder and finite thickness of the 2DES in reducing the
gap with respect to the numerically calculated one \cite{Boebinger85}.
However, the limited experimental range of variation of electron densities
and mobilities allowed only few data points for each incompressible state.
Detailed measurements of the activation gaps in the low-$B$ regime were
recently performed for filling factors $\nu =2/3$ and $2/5$ on high
mobility samples \cite{Schulze04}. The linearly vanishing gap close to
their spin-polarization quantum phase transition highlighted the
importance of the quasiparticle Zeeman energy in the spectrum at
relatively small magnetic fields.

In this paper we present a detailed analysis of the activation gap for the
paradigmatic FQH state at \emph{fixed} filling factor $\nu =1/3$ in a
\emph{very wide} range of \emph{purely perpendicular} magnetic field. For
the first time (to the best of our knowledge) we clearly observe a sharp
crossover between two regimes: a linear $B$-dependence for small magnetic
fields, due to spin-texture excitations, and a $\sqrt{B}$ dependence for
higher fields due to Coulomb interactions within the same spin channel.
The linear part directly yields the CF $g$-factor while the
disorder-induced reduction of the activation gap with respect to the ideal
clean case gives informations about the mobility edges of the CF Landau
levels (CFLL).\\

We now proceed to discuss the expectations of the free-CF theory on the
activation gap and subsequently present our experimental measurements with
the relative theoretical analysis.

In a FQH state at filling factor $\nu \equiv    n_{e}^{}\Phi_{0}^{}/B$
($n_{e}^{}$ the average electron density and $\Phi_{0}^{}=h/e$ the
magnetic flux quantum) the mismatch between the density of electrons and
of flux quanta induces a huge ground-state degeneracy at the
non-interacting level, making the many-body problem perturbatively
untreatable. Within the Chern-Simons picture \cite{Books}, CF are created
by attaching $\phi$ additional flux quanta to each electron ($\phi$ an
even integer), generating an additional magnetic field $b(\textbf{r})=\phi
\Phi _{0}n_{e}^{}(\textbf{r})$ ($n_{e}^{}(\textbf{r})$
 the local electron density) opposite to the external one, in order
to partially compensate for it and thereby reduce the degeneracy. This
gauge transformation depends only on the positions of the electrons and is
uncoupled to the Fermionic spin, thereby leaving the Zeeman splitting
unchanged. CFs are then subject to an effective magnetic field
$B^{*}(\textbf{r})=B-b(\textbf{r})$ that vanishes for $\nu =1/\phi$, on
the spatial average (mean field approximation). Near this filling factor
the cancellation is not exact and the residual $B^{*}=B\cdot (1-\phi\nu)$
yields CFLL with an effective cyclotron energy $\hbar \omecst\equiv \hbar
e B^{*}_{}/m_{}^{*}$ ($m_{}^{*}$ the CF effective mass). The CF filling
factor $p=n_{e}\Phi_{0}^{}/B^{*}_{}$ is related to the electronic one by
$\nu^{-1}_{} =p^{-1}_{}+\phi$, allowing the mean-field mapping of the
principal sequence of the electronic FQH states into integer QHE of CF. In
the following we will focus on the $\nu =1/3$ state (i.e. $p=1$) and
choose $\phi =2$, leading to $B^{*}_{}=B/3$.\\

The relevant energy scale involved in  electronic FQH states for GaAs
systems is the Coulomb repulsion $e^{2}_{}/\epsilon \ell\propto\sqrt{B}$
(with $\epsilon$ the dielectric constant and $\ell\equiv \sqrt{\hbar /e
B}$ the magnetic length), while the inter-LL cyclotron energy
$\hbar\omec\propto B$ is much larger for the interesting magnetic field
range. Thus, as long as spin effects are not concerned (or in the fully
polarized regime at high $B$), the CF cyclotron gap $\hbar\omecst$ is
expected to be proportional to $e^{2}_{}/\epsilon \ell$ (leading to a CF
effective mass $m^{*}_{}\propto\sqrt{B}$ \cite{HLR93,JainKamilla}) and
describes the electronic FQH gap. When spin is taken into account, each
CFLL is split into two sublevels separated by the Zeeman energy
$E_{\mathrm{Z}}^{}=g\muB B$, with $g$ the CF $g$-factor. The spectrum of
CFLL is therefore given by
\begin{equation}
\label{SCFLL} E_{n,s}(B)=\left(n+\frac{1}{2}\right)\hbar\omecst +s E_{\mathrm{Z}}^{}\quad ,
\end{equation}
with $n$ the CFLL index and $s=\pm 1/2$ \cite{ErosAnn}.

Within the non-interacting CF picture, the $\nu =1/3$ FQH state is mapped
onto the integer QHE at $p=1$, meaning that the zero-temperature ground
state is obtained by fully occupying the lowest CFLL $E_{0,1/2}^{}(B)$
(since the $g$-factor is negative in GaAs). In this way we easily recover
the full spin polarization of the $\nu =1/3$ state at $T=0$,
\emph{independent} on the value of $B$. However, due to the different
$B$-scaling of $\hbar\omecst$ and $E_{\mathrm{Z}}^{}$, the nature of the
excitation gap $\Delta_{\mathrm{id}}^{}$ (in the ideal clean case) is
different for low/high magnetic fields. In particular,
$\Delta_{\mathrm{id}}^{}=\min
[E_{0,-1/2}^{}(B),E_{1,1/2}^{}(B)]-E_{0,1/2}^{}(B)$ leading to
\begin{eqnarray}
\label{DeltaId} &&\Delta_{\mathrm{id}}^{}=E_{\mathrm{Z}}^{}\quad \; \mathrm{for}\; B<B_{\mathrm{c}}^{}\nonumber\\
&&\Delta_{\mathrm{id}}^{}=\hbar\omecst\quad \mathrm{for}\; B\ge B_{\mathrm{c}}^{} \quad ,
\end{eqnarray}
with $B_{\mathrm{c}}^{}$ such that
$E_{1,1/2}^{}(B_{\mathrm{c}}^{})=E_{0,-1/2}^{}(B_{\mathrm{c}}^{})$. Notice
that, in the low-$B$ regime, eventual deviations from the $\sqrt{B}$
dependence of $m^{*}_{}$, due to LL mixing, are immaterial for the
expectations of the excitation gap, which is anyway linear in $B$ and just
dependent on $g$. Thus, a measurement of the slope
$\partial_{B}^{}\Delta_{\mathrm{id}}^{}$ in the linear regime would
directly yield the CF $g$-factor. Similar arguments were used in
\cite{Schulze04} to extract $g$ at $\nu =2/3$ and $2/5$, but a
linearization of the CFLL energy close to $B_{\mathrm{c}}^{}$ was needed.
The $\nu =1/3$ case, in this sense, yields $g$ without any corrections due
to linearizations in the spectrum.

From the experimental point of view, the crossover from the linear to the
$\sqrt{B}$ scaling of the excitation gap was never clearly tested in a
\emph{single} sample until now. We succeeded in performing this experiment
due to the high-mobility of our 2DES and to the ability to modulate the
density in a very wide range keeping the filling factor constant while
varying the magnetic field.\\
Our 2DES is realized in an AlGaAs/GaAs heterostructure with a 70~nm thick
spacer from the $\delta$-doping layer. A metallic topgate enables us to
vary the electron density $n_{e}^{}$ between $2.0$ and $12.9 \cdot
10^{10}$ cm$^{-2}$ (see the inset of Fig.~1) with a zero field mobility
reaching $7 \cdot 10^{6}$ cm$^{2}_{}/V\, s$ at 40~mK. The filling factor
$\nu =1/3$ is then shifted from 2.5~T to 16.0~T with increasing density.
In Fig.~1, the longitudinal resistivity $\rho_{xx}$ vs $B$ is shown for
different densities at $T=40~mK$, the lowest temperature in our
experiment.
\begin{figure}[h]
\begin{center}
\includegraphics[width=7.5cm]{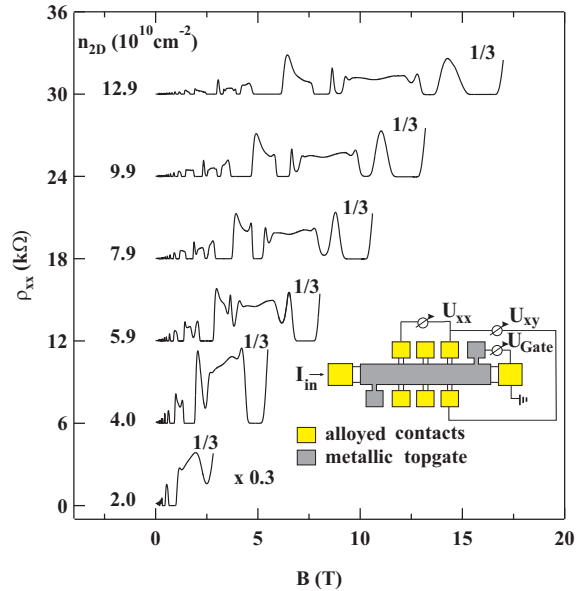}
\end{center}
\caption{Variation of the longitudinal resistivity $\rho_{xx}$ with the magnetic field at different electronic densities
$n_{e}^{}$. The different curves have been shifted for clarity. The inset shows the Hall bar geometry with a metallic
topgate.}\label{Fig1}
\end{figure}\\
To obtain the activation energy for the different magnetic fields we
investigate the temperature dependence of the resistivity minimum at $\nu
=1/3$. We extract the gap $\Delta$ out of the Arrhenius-plot data in
Fig.~2, using the activated resistance behavior $\rho_{xx} \propto
\exp({-\Delta/2T})$.
\begin{figure}[h]
\begin{center}\includegraphics[width=7.5cm]{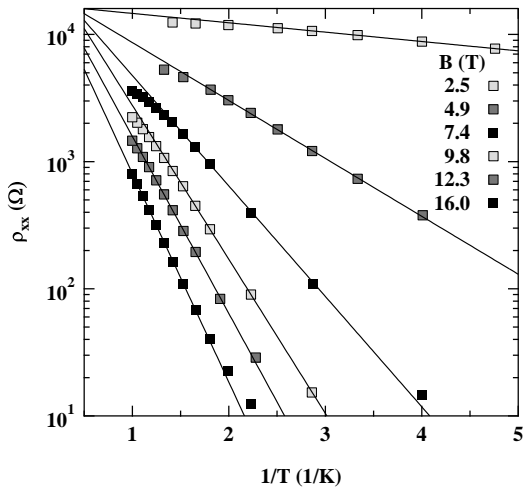}
\end{center}
\caption{Temperature dependence of the resistivity for filling factor $\nu=1/3$ at different perpendicular magnetic
fields. The solid lines are fits to the activated scaling $\rho_{xx} \propto \exp({-\Delta/2T})$.}
\label{Fig2}
\end{figure}
Finally, in Fig.~3 the measured activation energies $\Delta$ are plotted
versus the perpendicular magnetic field $B$.
\begin{figure}[h]
\begin{center}\includegraphics[width=7.5cm]{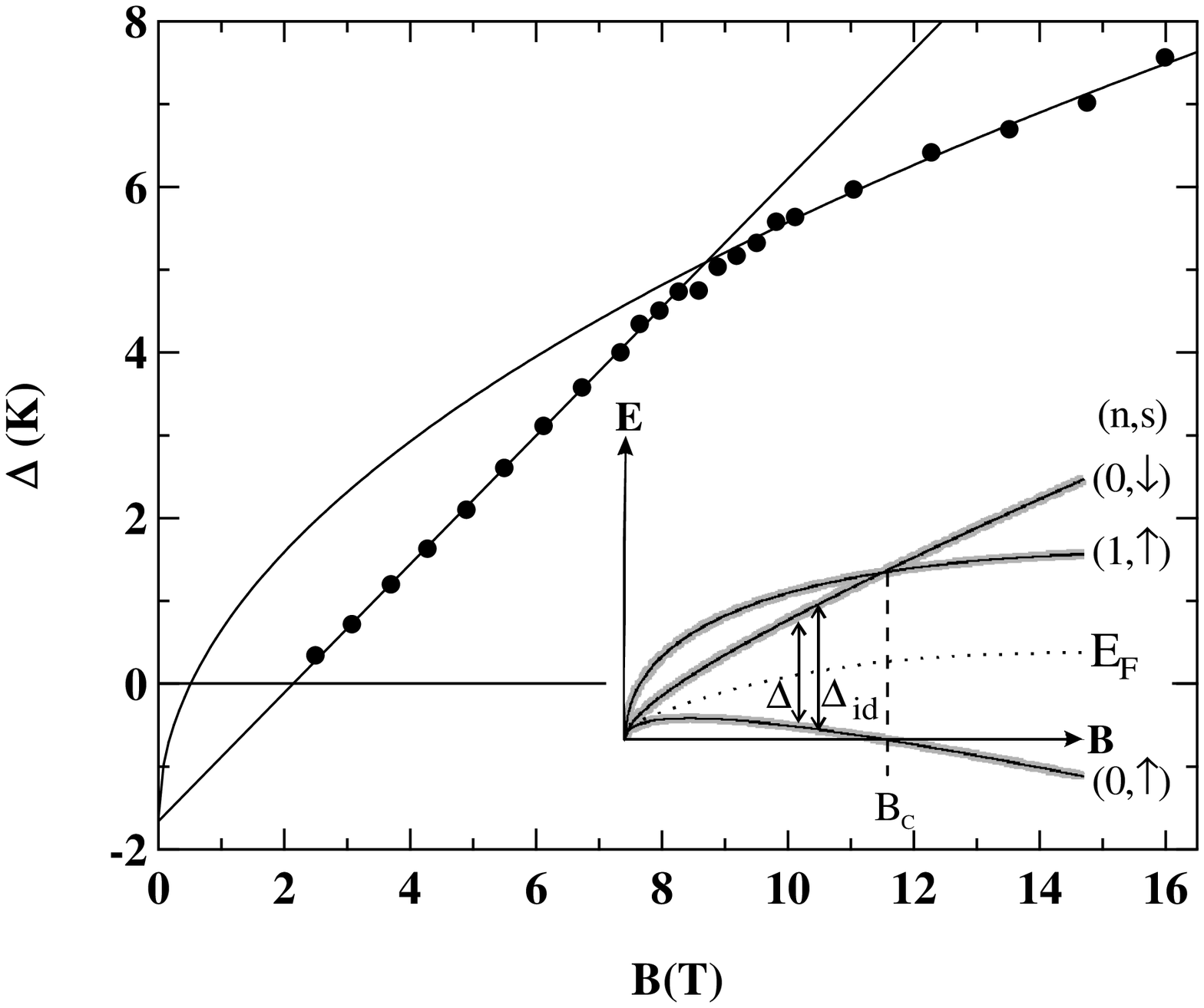}
\end{center}
\caption{The activation gap $\Delta$ at $\nu =1/3$ versus $B$. The linear and square
root scalings are clearly visible, as stressed by the two fit lines based on the free-CF model. A constant gap-reduction of $1.7\,$K has been included, yielding the negative intercept of the curves at $B=0$. The relevant disorder-broadened CF-Landau bands are depicted in the inset.}\label{Fig3}
\end{figure}

The first thing we notice is a sharp crossover between two different
scalings: a linear behavior at small fields and a $\sqrt{B}$ dependence
for $B$ larger than $B_{c}^{}=8.8\, \mathrm{T}$. This is, to the best
of our knowledge, the first time that such a feature is so clearly
observed, in agreement with the \emph{qualitative} expectations for
independent-CF.\\
A second important feature we observe is the disorder-induced reduction of $\Delta$, which vanishes at a finite
$B\simeq 2\,\mathrm{T}$. However, the functional dependence
of $\Delta$ on $B$ is the same as expected by the ideal theory, with no
appreciable corrections for the overall magnetic field range, suggesting a
\emph{magnetic field-independent reduction} of $\sim 1.7\,$K.\\
Finally we observe an additional gap reduction of about
$0.1\,\mathrm{K}$, smoothening the sharp transition close to $B_{c}^{}$.

Out of the slope $\partial_{B}^{}\Delta$ in the linear regime we directly
extract an effective $g$-factor at $\nu =1/3$ of $\left|{g}\right|=1.2$.
Previous measurements at fixed $\nu =2/3$ and $2/5$ \cite{Schulze04},
showed a strong dependence of $g$ on the filling factor, due to
interaction corrections \cite{AFS}, and suggested an extrapolated value at
$\nu =1/3$ of $\left|{g}\right|\simeq 0.44$ associated to a \emph{single}
spin-flip process. These arguments, however, neglect the residual CF interactions, in the
spirit of a quasiparticle picture of the original correlated problem. The
$\nu =1/3$ state is mapped onto a CF QH-ferromagnet at $p=1$ whose lowest
energy charged excitations in the low-B regime are expected to be smooth
spin-textures of the skyrmionic type \cite{Sondhi,Fertig,Kamilla}. Their
size is determined by the interplay between the Zeeman energy (favoring
small textures) and the exchange term of the residual CF-interactions
(favoring large skyrmions). \\
The linear part of the activation gap measured here seems to
suggest that the lowest charged excitations out of the CF QH-ferromagnet are
composite-skyrmions involving $\sim 3$ flipped spins, in agreement with
previous data \cite{Leadley}, highlighting the role of residual CF interactions.\\
Defining the CF mass as $m^{*}_{}=m_{0}^{}\alpha
\sqrt{B\mathrm{[T]}}$, with $m_{0}^{}$ the free electron mass in vacuum, we get $B_{c}^{}=4/(3 g\alpha)^{2}_{}$ and, in the
high-$B$ regime, $\Delta\mathrm{[K]}=-1.70+0.45\sqrt{B\mathrm{[T]}}/\alpha$ (having included the gap reduction). From both the measured values of the high-field gap and of $B_{c}^{}$ (which is not significantly affected by the disorder broadening of CFLL), we \emph{independently} get the \emph{same} CF mass parameter $\alpha =0.2$, in agreement with previous analysis \cite{Kukushkin99,ErosAnn}. This estimate once again shows the importance of finite thickness corrections \cite{Kukushkin00} to the quasiparticle effective parameters calculated from exactly 2D systems
\cite{ParkJain98,JainKamilla}.

In order to discuss the disorder-induced gap reduction,
we remind that $\Delta$ is rather the energy
difference between the first unoccupied mobility edge and the last occupied one. In the integer QHE, mobility edges are the energies at which
the localization length $\xi (E)$ equals the
typical sample size $L$. Within the scaling theory of the localization
\cite{Huckestein}, $\xi (E)$ diverges close to a LL center
as
\begin{equation}
\label{LocLength}
\xi (E)=\xi_{0}^{}\left|\frac{E-E_{c}^{}}{\Gamma_{\sigma}^{}}\right|^{-\gamma}_{}\quad ,
\end{equation}
with $E_{c}^{}$ the energy of the center of the Landau band,
$\Gamma_{\sigma}^{}$ the band-width (for a
disorder potential with correlation length $\sigma$) and $\gamma$ the
critical exponent. Finally, $\xi_{0}^{}$ can be extracted from the
localized regime in the LL tails and is related with the
correlation length $\sigma$, as a percolation picture suggests.
Thus, the mobility edges are $E_{c}^{}\pm
\Gamma_{\sigma}^{}\left(\xi_{0}^{}/L\right)^{1/\gamma}_{}$, yielding a gap
reduction $2\Gamma_{\sigma}^{}\left(\xi_{0}^{}/L\right)^{1/\gamma}_{}$
whose $B$-dependence is driven by
$\Gamma_{\sigma}^{}$.

In the case of CF, two disorder mechanisms broaden the
CFLLs into bands: a scalar random potential, mainly due to the
delta-doping outside the 2DES, and a random magnetic field (RMF)
scattering. The latter is due to fluctuations in the density which, via
the flux attachment,
produce fluctuations in the effective magnetic field felt by CF. It has been
shown \cite{Xie96} that the scaling law
(\ref{LocLength}) holds for CF as for electrons. The main difference is in the
scaling of $\Gamma_{\sigma}^{}$ due to the two scattering mechanisms.\\
The contribution to $\Gamma_{\sigma}^{}$ due to the scalar disorder,
$\Gamma_{\sigma}^{\mathrm{sc}}$, can be estimated within the
selfconsistent Born approximation \cite{AndoUemura} as
$\Gamma_{\sigma}^{\mathrm{sc}}=\Gamma_{}^{\mathrm{sc}}\left(1+\sigma^{2}_{}/\ell^{2}_{\mathrm{CF}}\right)^{-1/2}_{}$,
where $\Gamma_{}^{\mathrm{sc}}=2V_{0}^{}/(\sqrt{2\pi}\,
\ell^{}_{\mathrm{CF}})$ is the broadening due to a delta-correlated
disorder potential $V(\mathbf{r})$ with $\langle
V(\mathbf{r})V(\mathbf{r}^{\prime}_{})\rangle
=V_{0}^{2}\delta(\mathbf{r}-\mathbf{r}^{\prime}_{})$ and
$\ell^{}_{\mathrm{CF}}=\sqrt{3}\, \ell$ is the CF magnetic length. In our
case, the dominant impurity scattering is due to the $\delta$-doping
at a distance $d=70\,\mathrm{nm}$ from the 2DES, inducing a disorder with
a typical $\sigma\simeq d$. For all the magnetic fields in our experiment
($B>2\,\mathrm{T}$), $\ell^{}_{\mathrm{CF}} \ll \sigma$, yielding $\Gamma_{\sigma}^{\mathrm{sc}}\simeq
\Gamma_{}^{\mathrm{sc}}\ell^{}_{\mathrm{CF}}/\sigma=2V_{0}^{}/\left(\sqrt{2\pi}\,
\sigma\right)$. Since $V_{0}^{}$ is an intrinsic property of the
disorder potential independent of $B$, we deduce that
$\Gamma_{\sigma}^{\mathrm{sc}}$ is magnetic field independent,
leading to a \emph{constant} reduction of the
activation gap.\\
As far as the RMF contribution is concerned, the majority of results so
far have been obtained close to $\nu =1/\phi$, where $B^{*}_{}$ is small
(i.e. in the limit of high CFLL index $n$) \cite{EAltshuler}. Only
recently, the tails of the Landau bands of Fermions in a RMF with
non-vanishing average have been studied \cite{Mazzarello04}. In our case,
where $\ell^{}_{\mathrm{CF}}\ll \sigma$, the Fermionic states are
essentially cyclotron orbits in presence of the local (smoothly varying)
effective magnetic field $B^{*}_{}(\mathbf{r})=B^{*}_{}+\delta
B(\mathbf{r})$, with $\delta B(\mathbf{r})$ the RMF. The related typical
energy shift $\Gamma_{\sigma}^{\mathrm{RMF}}$ is the cyclotron energy
$\hbar e \delta B(\mathbf{r})/m^{*}$ and the RMF acts pretty much like a
scalar potential. Since $\delta B(\mathbf{r})$ is proportional to the
density fluctuations mainly driven by potential modulations, the relevant
$B$-dependence of the RMF broadening
is through $m^{*}_{}$, so that $\Gamma_{\sigma}^{\mathrm{RMF}}\propto B^{-1/2}_{}$.\\
A quantitative estimate of $\Gamma_{\sigma}^{\mathrm{RMF}}$ and
$\Gamma_{\sigma}^{\mathrm{sc}}$ is extremely difficult. However, the observed $B$-independent
gap reduction, particularly evident in the
sharp linear low-$B$ regime, suggests that $\Gamma_{\sigma}^{\mathrm{sc}}$ dominates over
$\Gamma_{\sigma}^{\mathrm{RMF}}$. Such a behavior may be understood via
the robust incompressibility of the $\nu =1/3$ state which strongly
suppresses the density fluctuations responsible for the RMF scattering.

The additional gap-reduction of about
$0.1\,\mathrm{K}$ close to $B_{c}^{}$ is probably due to an anti-crossing between the ($n=0, \downarrow $)
and ($n=1,\uparrow $) CFLLs, whose origin can be traced back to the spin-orbit coupling \cite{ErosAnn,Falko92} and whose
magnitude is consistent with the typical GaAs parameters.

In conclusion, we measured the activation gap of the $\nu =1/3$ FQH
ferromagnet in a wide range of the perpendicular magnetic field, observing
a sharp crossover between a spin-texture excitation regime at low $B$ and
a Coulomb-dominated one for large $B$. The residual CF interaction yields
skyrmion-like excitations at small $B$ involving $\sim 3$ flipped spins.
Furthermore, the constant gap reduction seems to indicate a dominant
scalar impurity scattering over the random magnetic field one.

We acknowledge discussions with R. Mazzarello, A. Shechter, F. Schulze-Wischeler and K. v. Klitzing and financial
support from the DFG via the priority program "Quantum Hall Systems".

\end{document}